# Exploration of the helimagnetic and skyrmion lattice phase diagram in $Cu_2OSeO_3$ using magneto-electric susceptibility

A. A. Omrani[1,2], J. S. White[1,3], K. Prša[1], I. Živković[4], H. Berger[5], A. Magrez[5], Ye-Hua Liu[6], J. H. Han[7,8], H. M. Rønnow [1*]

[1] Laboratory for Quantum Magnetism, Ecole Polytechnique Fédérale de Lausanne (EPFL),

1015 Lausanne, Switzerland

[2] Electrical Engineering Institute, Ecole Polytechnique Fédérale de Lausanne (EPFL), 1015 Lausanne, Switzerland

[3] Laboratory for Neutron Scattering, Paul Scherrer Institut, 5232 Villigen, Switzerland

[4] Institute of Physics, Bijenicka 46, HR-10000 Zagreb, Croatia

[5] Crystal Growth Facility, Ecole Polytechnique Fédérale de Lausanne (EPFL), 1015 Lausanne, Switzerland

[6] Zhejiang Institute of Modern Physics and Department of Physics, Zhejiang University, Hangzhou 310027, People's Republic of China

[7] Department of Physics and BK21 Physics Research Division, Sungkyunkwan University, Suwon 440-746, Korea

[8] Asia Pacific Center for Theoretical Physics, Pohang, Gyeongbuk 790-784, Korea

Abstract:

Using SQUID magnetometry techniques, we have studied the change in magnetization versus applied ac electric field, i.e. the magnetoelectric (ME) susceptibility dM/dE, in the chiral-lattice ME insulator $Cu_2OSeO_3$. Measurements of the dM/dE response provide a sensitive and efficient probe of the magnetic phase diagram, and we observe clearly distinct responses for the different magnetic phases, including the skyrmion lattice phase. By combining our results with theoretical calculation, we estimate quantitatively the ME coupling strength as $\lambda = 0.0146$ meV/(V/nm) in the conical phase. Our study demonstrates the ME susceptibility to be a powerful, sensitive and efficient technique for both characterizing and discovering new multiferroic materials and phases.

Multiferroic and magnetoelectric (ME) materials that display directly-coupled magnetic and electric properties may lie at the heart of new and efficient applications. Two intensely studied prototypical ME compounds with spiral order are $TbMnO_3$[1,2] and $Ni_3V_2O_8$[3], for which the microscopic mechanisms proposed to explain the generation of the electric polarization include the inverse Dzyaloshinksii-Moriya (DM)[4] and spin current[5] models, respectively.

Another exciting group of ME materials are chiral-lattice systems, since interactions that may promote symmetry-breaking magnetic order do not cancel when evaluated over the unit cell. The decisive role of non-centrosymmetry has been most clearly exemplified in itinerant $MnSi$[6,7], $FeGe$[8] and semi-conducting $Fe_{1-x}Co_xSi$[9]. In these compounds the principal phases are; 1) multiple q-domain helimagnetic order (helical phase) for $0<B<B_{c1}(T)$, 2) single-q helimagnetic order modulated along the field (conical phase) for $B_{c1}(T)<B<B_{c2}(T)$, and 3) a small phase pocket close to $T_N$ where a novel triple-q state described by three coupled helices ($\sum_{i=1}^{n} q_i = 0$) is stabilized. This latter phase is particularly interesting, since in MnSi the nano-sized ($\geq 15$ nm[10]) skyrmions can be coherently manipulated by the conduction electrons of an applied current[7], leading to both emergent electrodynamics[11], and promise for applications.

Most recently, the first Skyrmion lattice (SkL) phase in an insulating material was discovered in the chiral-lattice material $Cu_2OSeO_3$[12-14]. In direct analogy with the metal-like SkL compounds, $Cu_2OSeO_3$ also has the chiral-cubic $P2_13$ space-group, and the magnetic phase diagram is similarly composed of helical, conical and SkL phases[12,13]. The earlier proposed ferrimagnetic state in this compound exists for fields $B>B_{c2}(T)$[15-17]. The discovery at lower fields that $Cu_2OSeO_3$ displays the seemingly 'generic' magnetic phase diagram of a SkL compound is enthralling since a variety of studies show $Cu_2OSeO_3$ to display a ME coupling[12,15,18-21]. Indeed, the microscopic origin for the ME coupling is identified as caused by the *d-p* hybridization mechanism[21-24]. Most recently, emergent ME properties of the individual skyrmion particles were proposed[21,25], and demonstrated to exist experimentally[26]. $Cu_2OSeO_3$ represents a thus far unique opportunity for studying the electric field control of the magnetic properties in a ME SkL system.

The majority of the reported ME effects in materials are obtained from standard measurements of the electric polarization performed as a function of applied magnetic field and temperature[12,15,19,21]. Here we report measurements of the magnetoelectric susceptibility, which is the change in sample magnetization with ac electric field, to conduct a highly sensitive exploration of the ME effect across the entire magnetic phase diagram of $Cu_2OSeO_3$. While a similar approach has been applied previously, especially for exploring the ME effect in $Cr_2O_3$ [27], the level of detail our study on $Cu_2OSeO_3$ reveals across the rich helimagnetic phase diagram, combined with the quantitative estimate of the ME coupling strength obtained by comparing to theoretical calculations promotes this technique as highly efficient for discovering new multiferroics in general and new ME compounds with SkL phases in particular.

Single crystals of $Cu_2OSeO_3$ were grown by a standard chemical vapor transport method[16,20]. Our sample had a mass of 11.7 mg, a volume of $2\times2\times0.39$ mm$^3$, and was cut with the thinnest dimension along the [111] direction. Electrodes were created directly on the (111) crystal faces using silver paint. The sample was then mounted inside a vertical-field SQUID magnetometer, in which two different experimental geometries were studied: 1) E ∥ $\mu_0$H ∥ [111] and 2) E ∥ [111] with $\mu_0$H ∥ [1-10].

To measure the ME susceptibility, an ac electric field is applied to a single crystal sample and, in the presence of a simultaneous dc magnetic field, a SQUID magnetometer is used to monitor directly the associated change in sample magnetization. The change in the SQUID signal resulting from the applied ac electric field was recorded using a lock-in amplifier synchronized with the ac voltage generator.

The change in magnetization of the $Cu_2OSeO_3$ sample in configuration of E ∥ $\mu_0$H ∥ [111] is shown in Fig. 1(a). We observe that the response is linear in the electric field up to $7.7\times10^{-4}$ V/nm. Therefore, the gradient of the curve provides the change in magnetization as a function of the electric field, or the magnetoelectric susceptibility, $\chi_{ME}$, dM/dE for each magnetic field and temperature, and we can expect to model the phenomena with linear response theories and eg. Ginzburg-Landau models[28]. For the example of the (field-cooled) data at $\mu_0 H$=0.1 T and T=40 K, this variation is 6.6×10$^{-8}$ $\mu_B$/Cu per 1 V/m. In Fig. 1(b), the magnetic field-dependence of dM/dE is presented at various temperatures, covering the helical, conical and ferrimagnetic phases. Salient features include the linear tendency of dM/dE of different slopes within the conical phase, and the drop of the signal for fields B>B$_{c2}$(T) in the ferrimagnetic phase.

Since the SkL phase is reported to exist in the approximate temperature range of 56-58 K, in Fig. 2 we show high precision magnetic field-dependent measurements of the ac magnetoelectric and magnetic susceptibilities, and the dc magnetization at T=57 K for the two different experimental configurations E || $\mu_0$ H || [111] (Fig. 2 (a)-(c)) and E || [111] with $\mu_0$ H || [1-10] (Fig. 2 (d)-(f)). The magnetoelectric susceptibility is seen to be a particularly revealing probe of the magnetic phase diagram; a series of sharp peaks and dips are observed in both the real and imaginary parts that give clear evidence for magnetic transitions. A remarkable feature of these data is the high precision at which these transitions may be determined, compared to the corresponding kink-like features in the ac magnetic susceptibility (Fig. 2 (b) and (e)), and only small wiggles seen in the dc magnetization (Fig. 2 (c) and (f)).

For both magnetic field geometries, the magnetic field-dependence of the ME susceptibility can be divided into three main parts. Firstly, the value of dM/dE remains very small in the helical phase and seemingly constant in SkL phases. Secondly, dM/dE depends linearly on the applied magnetic field within the conical phase on both sides of the Skyrmion phase. This is most easily seen in the high field regime, where also the maximum signal in dM/dE is observed. Thirdly, both the lower and upper field borders between the SkL and conical phases are characterized by strong peaks and dips in both the real and imaginary parts of dM/dE. The observed peaks and dips for the transition into and out of the SkL phase reflect nonlinear responses occurring at the phase boundaries. This observation contrasts the behavior seen for the transitions at both B$_{c1}$ (helical-conical transition), and B$_{c2}$ (conical-ferrimagnetic transition), where only the real part of the magnetoelectric susceptibility shows steps. Extra measurements were carried out at T<56 K where no SkL phase is expected, and only the transitions at $B_{c1}(T)$ and $B_{c2}(T)$ are observed. This confirms that the extra peaks and dips observed at T=57 K may be assigned to transitions on the borders of the SkL phase.

The magnetic field- and temperature-dependence of the real and imaginary parts of dM/dE, and the dc magnetization are shown for the E ∥ $\mu_0$ H ∥ [111] geometry in Fig. 3 (a)-(c). By tracking the peak and dip features observed in temperature-scans of dM/dE, the main magnetic phases are easily identified (particularly in the real part of dM/dE), and agree well with the phase diagram determined with alternative methods[12,13]. The data shown in Fig. 3 (b) also indicate that the large peaks and dips in the imaginary part of dM/dE occur at the SkL phase boundary. In Fig. 3(d) the first magnetic phase diagram constructed by using the ac magnetoelectric susceptibility technique is presented. Only weak traces of these transitions are seen in the phase diagram produced by dc magnetisation (Fig. 3 (c)). Above 58 K the continuous decay of the signal with increasing temperature indicates a regime of short range order. The field dependence in this regime is linear with slope of similar magnitude as in the conical phase and there are no signs of cross-overs. We therefore conclude that the critical fluctuations are short range correlations of the conical order type.

Next we develop a theoretical framework for calculating $\chi_{ME}$. We consider the effective Hamiltonian $H = H_{HDM} + H_{ME}$[29]. We approximate the microscopic Hamiltonian of the system with a simplified one where one effective magnetic moment $S_i$ represents the total unit-cell moment in $Cu_2OSeO_3$. The first term includes Heisenberg, Dzyaloshinskii-Moriya (DM) spin-spin interaction and the Zeeman term[25]. The second term $H_{ME} = -\sum_i \mathbf{P}_i \cdot \mathbf{E}_i$ includes the ME coupling where the local electric dipole moment per unit cell is coupled to the spin configuration according to[21,25]

$$\mathbf{P}_i = \lambda(S_i^y S_i^z, S_i^z S_i^x, S_i^x S_i^y) \tag{1}$$

The coupling constant λ here represents the strength of the ME coupling between the effective unit cell moments and the electric field. If the direction of the applied electric field is $\hat{e}$ and we are interested in the magnetization along $\hat{m}$, the ME susceptibility is obtained from

$$\chi_{ME} = \frac{\partial(\boldsymbol{M}.\hat{m})}{\partial E} = \frac{\lambda g \mu_B}{NT}\left(\left\langle \left[\sum_i \boldsymbol{S}_i \cdot \hat{m}\right]\left[\sum_i \boldsymbol{P}_i \cdot \hat{e}\right]\right\rangle - \left\langle \left[\sum_i \boldsymbol{S}_i \cdot \hat{m}\right]\right\rangle\left\langle \left[\sum_i \boldsymbol{P}_i \cdot \hat{e}\right]\right\rangle\right) \quad (2)$$

where the magnetization is $\boldsymbol{M} = g\mu_B \langle\sum_i \boldsymbol{S}_i\rangle/N$, and $N$ is the number of unit cells (in our simulation $N = 12^3$) and T is the temperature. The averages $\langle...\rangle$ are performed by means of Monte-Carlo simulation. For the case of B || E || [1 1 1], we can choose $\hat{e} = \hat{m} = (111)/\sqrt{3}$. The results of such calculations are shown in Fig. 4(b).

Additionally, a Ginzburg-Landau (GL) approach is used to derive the linear ME response in the conical phase. In the same geometry as introduced above, the full GL free energy density has the form:

$$F = \frac{J}{2a}(\nabla\boldsymbol{\mu})^2 + \frac{D}{a^2}\boldsymbol{\mu}\cdot(\nabla\times\boldsymbol{\mu}) - n|S|\frac{B}{a^3}(\mu_x + \mu_y + \mu_z)/\sqrt{3} - \lambda\frac{E}{a^3}(\mu_y\mu_z + \mu_z\mu_x + \mu_x\mu_y)/\sqrt{3} \quad (3)$$

where $\boldsymbol{\mu}$ is the sample magnetization J, D, n and a are the Heisenberg, DM coupling energies, number of copper sites in the unit cell and lattice constant, respectively. After a rotation of both real and spin coordinates[25], the [1 1 1] direction lies along the z-axis in the rotated frame and Eq. (3) becomes

$$\frac{F}{8JK^2} = (\nabla\boldsymbol{\mu})^2 + \boldsymbol{\mu}\cdot(\nabla\times\boldsymbol{\mu}) - \beta\mu_z - \frac{1}{2}\varepsilon\mu_z^2 \quad (4)$$

where $\beta = n|S|B/(8JK^2a^3)$ and $\varepsilon = \sqrt{3}\lambda E/(8JK^2a^3)$, and the space coordinates are re-scaled as $r \to r/(4\kappa)$ with $\kappa = D/(2J)$. The dimensionless free energy in Eq. (4) facilitates the following discussion. The right-handed conical unit-length spin configuration in a 3D material which is compatible with D > 0 is described by

$$\boldsymbol{\mu}(x,y,z) = [\sin(\theta)\cos(qz), \sin(\theta)\sin(qz), \cos(\theta)] \quad (5)$$

where $|\boldsymbol{\mu}| = 1$, $\theta$ is the conical angle, and (0,0,q) is the conical modulation vector. By inserting (5) into the energy functional of Eq. (4), and minimizing with respect to both q and θ, we get $q_0 = ½$ and $\cos(\theta_0) = \frac{2\beta}{1-2\varepsilon}$. The derivative with respect to ε becomes $\frac{4\beta}{(1-2\varepsilon)^2}$, and hence by considering average unit cell magnetization as $|\mathbf{M}| = gn\mu_B|S|$ with same spin configuration as $\boldsymbol{\mu}$ the derivative with respect to ε forms as:

$$\frac{\partial(M_z)}{\partial \varepsilon} = (gn\mu_B|S|)\frac{4\beta}{(1-2\varepsilon)^2} \tag{6}$$

which depends linearly on the magnetic field. The final expression for the low electric-field limit of the ME susceptibility containing material parameters in the conical phase can be written as:

$$\chi_{ME} = \frac{\partial(M_z)}{\partial E} = \frac{\partial M_z}{\partial \varepsilon}\frac{\partial \varepsilon}{\partial E} = 4\sqrt{3}(gn\mu_B|S|)^2 \frac{\lambda}{(8JK^2)^2} B \tag{7}$$

We now discuss how our experiments compare to the expectations of the theoretical estimates. Fig. 4(a) shows a magnetic field scan of dM/dE done at 54 K which, as seen from Fig. 3, is a temperature where no SkL phase exists. At low fields, only a very small signal is observed in the helical phase. Upon increasing the field, a jump is observed in dM/dE at the transition into the conical phase, where after we observe a linearly increasing signal until the sharp fall upon entering the ferrimagnetic phase. This behaviour is in good qualitative agreement with the results of Monte-Carlo simulations (e.q. 2) presented in Fig. 4(b). Furthermore, by using Eq. 7 derived in the GL approach, we can estimate quantitatively the size of the effective ME coupling parameter λ. The slope of dM/dE extracted in the conical phase at 54 K is $1.58\times10^{-4}$ ($\mu_B$/Cu)(V/nm)$^{-1}$(Oe)$^{-1}$. For $Cu_2OSeO_3$ the effective Heisenberg coupling between unit cell moments is chosen to be J = 3.4 meV, which reproduces the correct ordering temperature. The ratio κ = D/2Ja = π/l is determined from the wavelength l = 630 Å [13,26] of the magnetic helix relative to the lattice constant a = 8.9 Å [30]. With $|S| = \langle S_z \rangle$ in the ferrimagnetic phase determined to be 3.52 $\mu_B$/unit cell at 54 K, we find $\lambda$ = 0.0146 meV/(V/nm) = 2.34×10$^{-33}$ J/(V/m). This value of the ME coupling leads to local electric dipole moment of unit cell $P$ = 7.216×10$^{-27}$ $\mu$C.m based on eq. 1 or macroscopic polarization $p$ = 10.2 $\mu$C/m$^2$ which is of the same order of magnitude as reported by Seki et. al[12].

The observed behaviour in dM/dE when passing through the SkL phase at 57 K is more complicated (Fig. 4(c)). We interpret the signal as a contribution of a piece-wise linear response and sharp non-linear peaks at the transitions. Due to the strong non-linear peaks, the exact field dependence of the response in the SkL phase cannot be determined precisely and is thus presented as a shaded green area in Fig. 4(d). The sharp peaks are ascribed to the non-linear response related to the first order transitions separating the conical and SkL phases. The imaginary components of the peaks have opposite sign to the real part. A possible explanation is that varying the magnetic field places the system in a higher energy out-of-equilibrium state, whereby each E-field ac cycle releases, rather than absorbs, energy. The observation that this non-linear effect occurs exclusively around the SkL phase borders could indicate near-degeneracy of many quasi-protected non-perfect SkL configurations that couple strongly to the E-field. In turn, this provides exciting prospects for the future E-field control of individual skyrmions.

In conclusion, we have presented a ME susceptibility study of the phase diagram and ME coupling in Cu$_2$OSeO$_3$. By exploiting the superior sensitivity of a SQUID magnetometer, magnetization changes as small as 10$^{-3}$ emu.nm/V are detected for a 10 Hz and 5 V driving ac electric field, and allow the efficient exploration and characterization of the ME coupling across the helimagnetic phase diagram of the chiral-lattice ME Cu$_2$OSeO$_3$. Furthermore, first principle calculations of the ME susceptibility provide a quantitative analysis of the data, as exemplified by the extraction of the ME coupling parameter $\lambda$ = 0.0146 meV/(V/nm). This work demonstrates ME susceptibility measurements to be a technique of choice for studying the general properties of ME compounds with rich magnetic phase diagrams, and opens the door for new investigations of multiferroic skyrmions, most notably their manipulation by electric field.

**Acknowledgements**

We gratefully acknowledge financial support from the Swiss National Science Foundation, MaNEP and the European Research Council.

Figure 1:

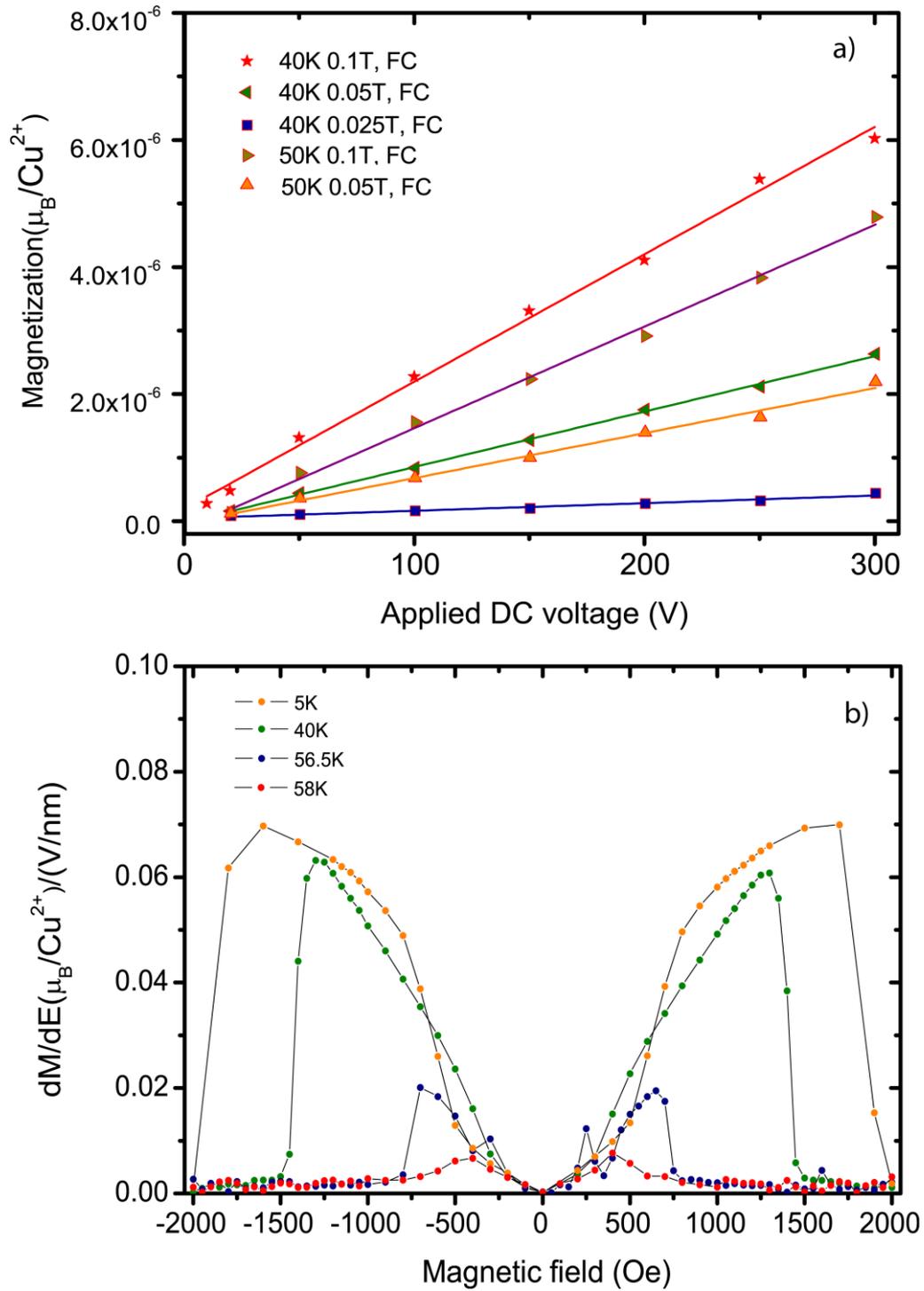

**FIG. 1.** **(a)** Magnetoelectric signal as function of constant electric field applied along [111] for various magnetic field and temperature conditions, **(b)** magnetic field scans of the ac magnetoelectric susceptibility measured at different temperatures (no demagnetization correction is applied here).

Figure 2:

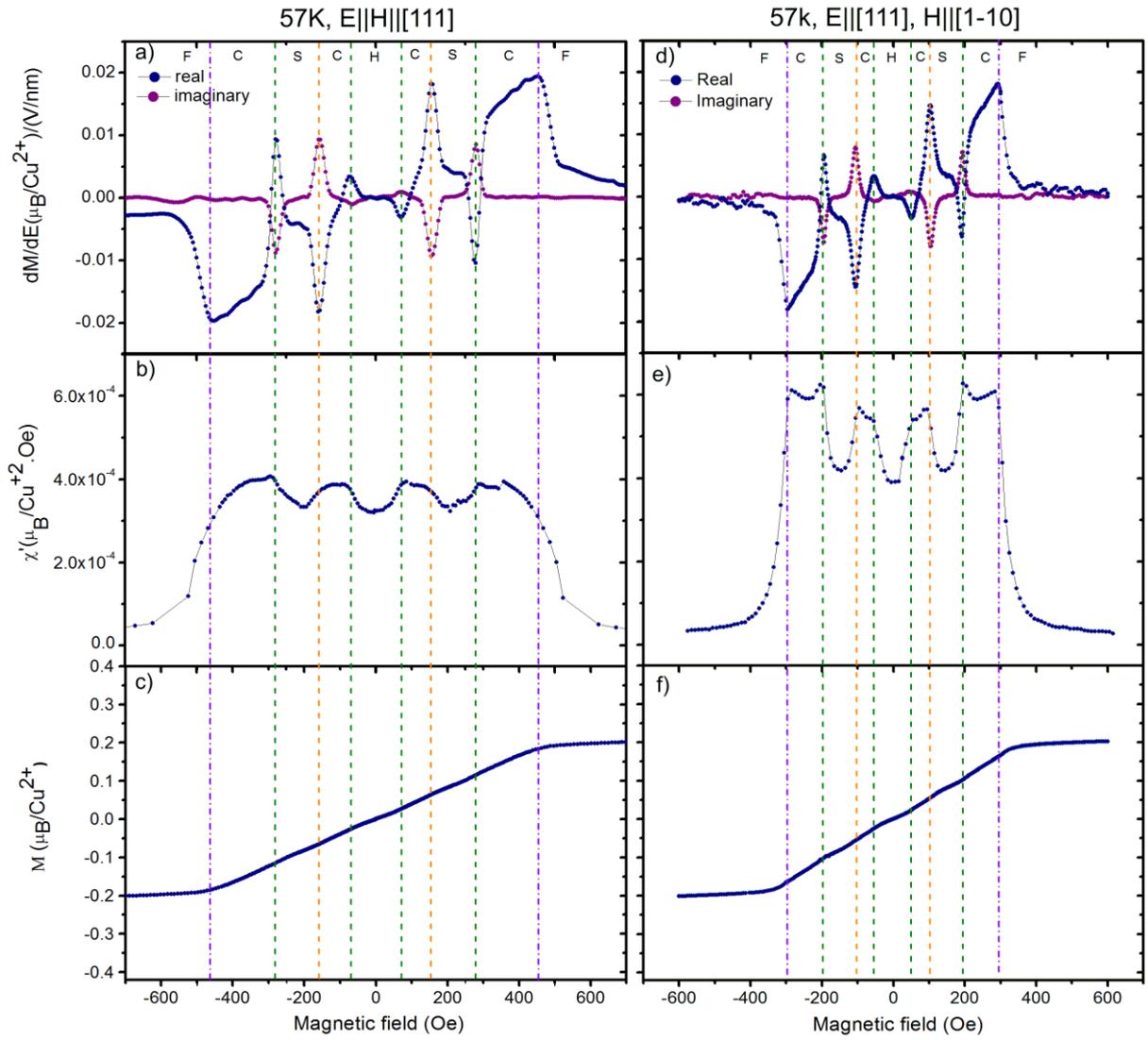

**FIG. 2.** The magnetic field-dependence of: **(a), (d)** the ac magnetoelectric susceptibility, **(b), (e)** ac magnetic susceptibility, and **(c), (f)** dc magnetization for E ∥ $\mu_0$H ∥ [111] **(a)**-**(c)**, and E ∥ [111] with $\mu_0$H ∥ [1-10] **(d)**-**(f)**. In the latter crystal orientation, due to the small area exposed to magnetic field, no demagnetization correction has been made. All measurements of dM/dE were done using a 10 Hz, 5 V ac voltage. The letters F, C, S and H denote the ferrimagnetic, conical, skyrmion and helical phases, respectively.

Figure 3:

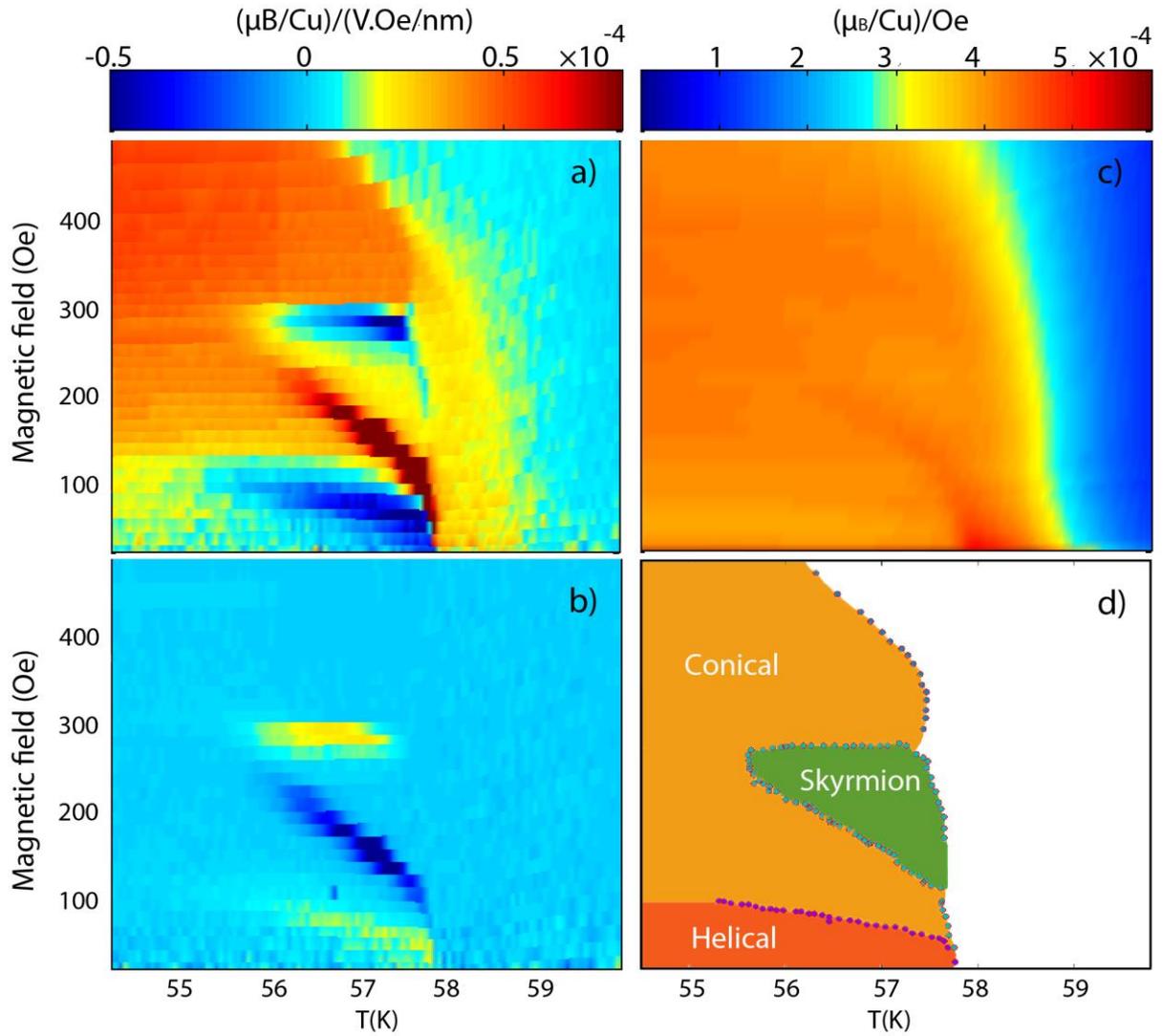

**FIG. 3.** For the E ∥ $\mu_0$H ∥ [111] geometry, magnetic phase diagrams constructed using **(a)** the real, and **(b)** imaginary parts of the magnetoelectric susceptibility, and **(c)** the dc magnetization. These diagrams were constructed using temperature scans (warming) after the sample was field-cooled from 70 K. In **(d)** we show the portion of the magnetic phase diagram near the ordering temperature extracted from the real part of the temperature scans signals.

Figure 4:

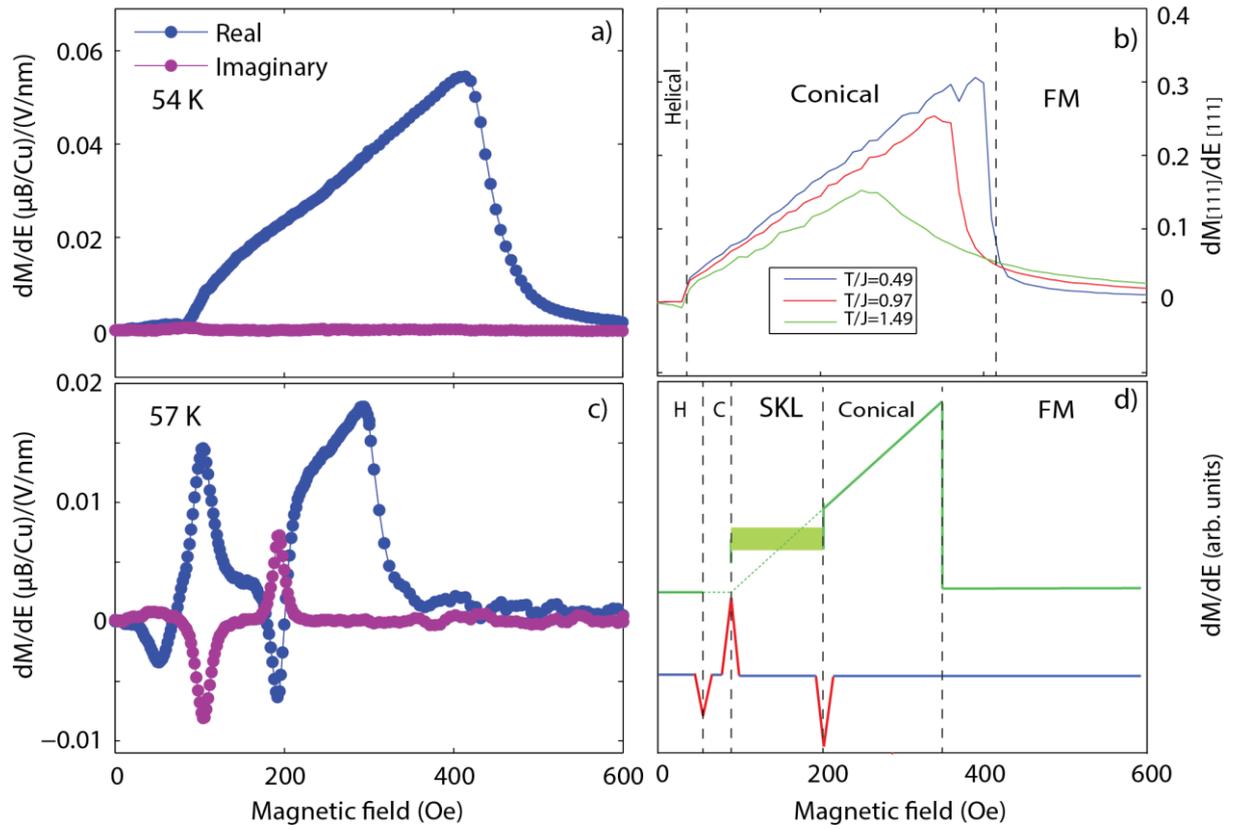

**FIG. 4.** Real and imaginary part of the magnetoelectric response for the E ∥ [111] with $\mu_0 H$ ∥ [1-10] geometry at 54 K **(a)** and 57 K **(c)**, respectively. Part **(b)** represents the simulation results of a 3D lattice hosting helical, conical and ferrimagnetic phases in the E ∥ $\mu_0 H$ ∥ [111] geometry. In **(d)** a schematic of the piece-wise linear behavior of dM/dE in the conical phase, and also including peaks and dips on each side of the SkL phase.